\documentclass[10pt,showkeys,showpacs,amsmath,amssymb,prl]{revtex4}
\usepackage[english]{babel}
\usepackage[T2A]{fontenc}
\usepackage[cp866]{inputenc}
\usepackage[dvips]{graphicx}
\usepackage{amsmath,euscript,amssymb}
\newcommand\bea{\begin{eqnarray}}
\newcommand\eea{\end{eqnarray}}
\newcommand\be{\begin{equation}}
\newcommand\ee{\end{equation}}

\def\gsim{\mathrel{\raise.3ex\hbox{$>$\kern-.75em\lower1ex\hbox{$\sim$}}}}
\def\lsim{\mathrel{\raise.3ex\hbox{$<$\kern-.75em\lower1ex\hbox{$\sim$}}}}

\begin{document}
\title{Relativistic Universe Scenario
 }
\author{V.N. Pervushin}
\email{pervushin@theor.jinr.ru}
\affiliation{Bogoliubov Laboratory
of Theoretical Physics, Joint Institute for Nuclear Research, 141980
Dubna, Russia}

\date{\today}
\pacs{ 95.30.Sf, 98.80.-k, 98.80.Es} \keywords{General Relativity and Gravitation, Cosmology, Observational Cosmology}
\begin{abstract}
 The luminocity-distance -- redshift SN Ia data and  the Cosmic Microwave
 Background power spectrum are discussed in the context of the
  results and ideas of the eminent Russian physicist-theorist
 N.A. Chernikov. His results include the
 Boltzmann-Chernikov distribution in
 the kinetic theory of relativistic gas, conformal invariant theory of scalar field, and
  the vacuum cosmological
 creation  of particles. We use these results for  explanation of the origin of matter content of the Universe
  in  modern relativistic physics.
\end{abstract}
\maketitle
 \centerline{\it The talk presented at the BLTP Seminar dedicated to the memory of Professor N.A.Chernikov,}
\centerline{Dubna, Russia, April 17, 2008}

{\small In collaboration with A.B. Arbuzov, B.M. Barbashov,
A. Borowiec ({\it Institute
 of Theoretical Physics, University of
 Wroc\l aw}), S.A. Shuvalov ({\it Russian Peoples Friendship
University}), and A.F. Zakharov ({\it Institute of Theoretical and
Experimental Physics, Moscow})}

\section{Relativistic Physics}

The best introduction to  relativistic physics was given by David
 Hilbert in his G\"ottingen talk  ``Die Grundlangen der Physik'' \cite{H-1915},
 where the Einstein equations \cite{einsh-16}
in  General Relativity were derived by the variational principle.
In accord with the G\"ottingen ``Foundations of Physics'', differences
of the trajectory of a relativistic particle
\be\label{rel-1}
 {{X}^{(\mu)}_A- {X}^{(\mu)}_B}={V}^{(\mu)}\cdot s~~ \Big |^{ds=N(x^0)dx^0}_{V^{(\mu)}V_{(\mu)}=c^2}
 \ee
from the Galilei  trajectory of a nonrelativistic particle
 ${{\vec X}_A- {\vec X}_B}={\vec V}\cdot t$
 are  the 4-dimensional Minkowski space of events (instead of 3-dimensional space)
  and tree ``times'' (instead of  the single    absolute Newton time $t$) required for
  complete description of the motion of a relativistic particle:


 i. a relative ``time-variable'' $X^{(0)}$ in the space of events as an object of
the Poincar\'e  transformations of a frame of reference, measured by an observer in his frame,

 ii. an absolute ``time-interval'' $s$ at the particle trajectory in
space of events, given as the one-dimensional Riemannian manifold, and


  iii. an unmeasurable ``time-coordinate'' $x^0$ in this manifold as an object of its
reparametrizations $x^0\to \widetilde{x}^0=\widetilde{x}^0(x^0)$.


    The reparametrization group in the  Riemannian manifold $ds=N(x^0)dx^0$
     is a new element   in comparison with the frame group in the classical mechanics. In particular,
    the Hilbert variation \cite{H-1915} with respect to the metric component $N(x^0)$ leads to the velocity constraint
 of the type of the Lobachevsky space $V^{(\mu)}V_{(\mu)}=V^2_{(0)}-V^2_{(k)}=c^2$.
 This constraint is a complete analog of the Einstein equations in GR. The relativistic postulates
 (\ref{rel-1}) have firm evidences  beginning from
  numerous experimental facts revealed in 19th century, including  the first dynamo,
 the concepts of the field nature of electro-magnetic phenomena formulated by Faraday who gave an idea of unification of all interactions,  finishing with Maxwell's equations, their interpretations by Einstein, Lorentz, and the Poincar\'e group of these equations.

 Irreducible unitary representations of the Poincar\'e group  constructed by Wigner in 1939
 \cite{Wigner-39} as the basis of  quantum field theory \cite{Logunov}
 mean the relativistic
classification of physical states marked by their masses and spins.
The difference between the Poincar\'e group classification of physical
states and the classification of fields with respect to the Lorentz
subgroup (given in the  Lobachevsky space of velocities) determines the set of
gauge constraints. Using these constraints and the second N\"other  theorem
Ogievetsky and Polubarinov \cite{op-65} could restore at the quantum level
all field interactions (including QED, GR, YM, QCD) besides
the Higgs potential and the Penrose-Chernikov-Tagirov action \cite{pct,JINR-1}.

The problem of  finite amplitudes of weak quantum transitions between physical
states in GR in the 60s was not so dramatized, like at the  present time.
  Abdus Salam constructed a finite S-matrix element in GR  using for regularization
 the superpropagator method developed by M.K. Volkov \cite{MKV}. In the 60s the absence of the
 standard renormalization scheme  for GR was not yet considered as an argument to ignore
 the Poincar\'e group classification of physical states in GR,
in contrast to the recent  models
 accepted for description of recent observational data where this  classification of physical states
 is ignored.

In particular,
the Poincar\'e group classification of physical states~\cite{Wigner-39} does not
include the scalar metric component that is used now in
the accepted $\Lambda$CDM cosmological model~\cite{WMAP01,Giovannini}  for  explanation of the CMB
anisotropy. Recall that this
classification is based on the vacuum postulate about the existence of a state with a minimal energy.
 The vacuum postulate removes  free dynamics of all negative energy components
 including the dynamics of scalar metric component used in the $\Lambda$CDM model as
the origin of the CMB acoustic peaks.

The paper is devoted to the application of  results  by
 N.A.~Chernikov for solution of these topical problems of  modern Cosmology
   including  the CMB power spectrum,  SN data, and the origin of matter in the Universe.

Nikolay Aleksandrovich Chernikov is
the outstanding Russian theoretical physicist (16.12.1928 --  17.04.2007) \cite{CP-08}. His first steps in clarifying
the role of geometry in relativistic physics and   stochastic relativistic motion \cite{CH-57} were supported by V.A. Fock. Chernikov  formulated
the kinetic theory of relativistic gas, including the Boltzmann-Chernikov distribution function \cite{CH-63,Muller-2007}. Together with his coauthors Chernikov supposed and developed
  the conformal invariant theory of scalar field \cite{JINR-1}. The idea
 of the vacuum cosmological
 creation  supposed by Chernikov  and his pupils \cite{JINR-1,JINR-2} independently from
 L. Parker \cite{Parker-69} was very highly estimated by Ya.B. Zel'dovich, A.A. Starobinsky, and A.A. Grib
 (see \cite{Yaf-71,grib80}).

 We show how these Chernikov's results
help us to solve the problems of description of the SN data and CMB anisotropy without violation of  principles of quantum relativistic
physics associated with  Poincar\'e, Einstein, Hilbert, Wigner, Dirac and other scientists.

\section{Model of a  Relativistic Universe}
The modern cosmological approaches
 including the widely accepted $\Lambda$CDM model~\cite{Giovannini,lif,MFB}
 are based on the General Relativity (GR) supplemented by the Standard Model (SM)~\cite{weak}
and by an additional scalar field $Q$  governing the Universe evolution \cite{Guth_92}:
 \be\label{1-1}
 S_{\rm U}[F]=\int\limits_{ }^{ }d^4x\sqrt{-g}\left[-\frac{1}{6}R(g)
 +{\cal L}_{\rm SM}(F)+\partial_\mu Q\partial^\mu Q -\textsf{V}_{\rm U}(Q)\right]
 \ee
in the Riemannian space with the interval $ds^2=g_{\mu\nu}dx^\mu
dx^\nu$, where $F=f,v,\phi$ is a set of the SM fields, fermion $f$,
vector $v$, and scalar $\phi$. Throughout the paper we  use the
units
 $ 
 \hbar=c=M_{\rm Planck}\sqrt{{3}/({8\pi}})=1
 $.

 In accord with the so-called cosmological principle introduced by Einstein and Friedmann
 \cite{Einstein_17,fried_22},
 all local scalar characteristics of the Universe evolution
 averaged over a large coordinate volume $V_0=\int d^3x $, \textit{i.e.} zeroth harmonics,
 \bea\label{z-s1-1}
 \log a\equiv\frac{1}{6V_0
 }\int d^3x \log |g^{(3)}|,~~~~\langle
 \phi\rangle\equiv\frac{1}{V_0
 }\int d^3x \phi,~~~~~\langle
 Q\rangle\equiv\frac{1}{V_0
 }\int d^3x \,Q
 \eea
depend only on the world time $dt =a(\eta)d\eta$ in the conformal-flat  interval
 \be \label{cti-2}
 ds^2=a^2(\eta)[(d\eta)^2-\sum_{j=1}^3(dx^j)^2]\equiv a^2(\eta) \widetilde{ds}^2,
 \ee
 where $d\eta=N_0(x^0)dx^0$ is the diffeo-invariant conformal time, and
$N_0(x^0)=\langle \sqrt{-\widetilde{g}}\,\widetilde{g}^{00}\rangle^{-1}$
 is the diffeo-variant global lapse function arising in the second term of the action
 \bea\label{eta-1}
 S_{\rm U}[F]\Big |_{g=a^2\widetilde{g},f=a^{-3/2}\widetilde{f},\phi=a^{-1}\widetilde{\phi}}\equiv
 \widetilde{S}_{\rm U}[a|\widetilde{F}]\Big |_{\widetilde{F}=(\widetilde{g},\widetilde{f},\widetilde{\phi})}
 -V_0\int\limits_{\eta=0}^{\eta_0} d\eta a'^2, ~~~~~a'=da/d\eta,
 \eea
that emerges after the conformal transformations of fields in action
(\ref{1-1}) \cite{Barbashov_06,B_06,Bog,Glinka_08}. The choice of interval
(\ref{cti-2}) in action  (\ref{eta-1}) is known as the cosmological
approximation of GR~\cite{fried_22,Narlikar}, where the inverse
scale-factor,  $a^{-1}(\eta)=z+1$, is treated as the redshift of an
atomic spectrum of an observable cosmic object being at the
coordinate distance $r=\eta_0-\eta$; here $\eta_0$ is the
present-day value of the conformal time distinguished by
$a(\eta_0)=1$, and $\eta$ is the instance of the photon emission by
an atom in the given object. In this approximation, the Hilbert variation of
the action with respect to the diffeo-variant global lapse function $N_0(x^0)$
gives the diffeo-invariant cosmological equation treated as the energy constraint
 \cite{H-1915}
 \bea\label{free-c1a}
 a'^2&=&\widetilde{\rho}_{\rm U} (a)\equiv\frac{1}{V_0}\frac{\delta \widetilde{S}_{\rm U}[a|\widetilde{F}]}{\delta N_0(x^0)}=
a^2[\langle \phi' \rangle^2+\langle Q' \rangle^2]+
 a^4\left[\textsf{V}_{\rm Higgs}(\langle\phi\rangle)+\textsf{V}_U(\langle Q\rangle)\right]+
  \dfrac{\textsf{H}_{\rm QFT}(\widetilde{F})}{V_0}\, ,
 \eea
  where $\widetilde{\rho}_{\rm U} (a)$ is the conformal density and the field energy
  $\textsf{H}_{\rm QFT}(\widetilde{F})$
  is the standard component of the  energy-momentum tensor of all fields $\widetilde{F}$
  \be\label{chh-2e}
 \textsf{H}_{\rm QFT}[\widetilde{F}]
 =\sum\limits_{F, \textbf{l},\textbf{l}^2\not
 =0}^{}\widetilde{n}_{\widetilde{F},\textbf{l}}
 \omega_{F,\textbf{l}}(a)+\textsf{H}_{\rm int}[\widetilde{F}]
 \ee
   in any QFT model in the flat space-time with the interval $\widetilde{ds}^2$ given by Eq. (\ref{cti-2}), where $\omega_{F,\textbf{l}}(a)=\sqrt{{\bf k}^2+\widetilde{m}^2_{\rm F}}$ is  one-particle energy
   with the particle masses
  $\widetilde{m}_{\rm F}$   defined by the  zero mode of the Higgs field $\widetilde{m}_{\rm F}= g_{\rm F}a(\eta)\langle\phi\rangle(\eta)=
   g_{\rm F}\langle\widetilde{\phi}\rangle(\eta)$
   with the corresponding coupling constant $g_{\rm F}$, and
${\widetilde{n}_{F,\textbf{l}}}=[\widetilde{F}^+_\textbf{l}~\widetilde{F}^-_\textbf{-l}\pm \widetilde{F}^-_\textbf{-l}~\widetilde{F}^+_\textbf{l}]/2$
(with a positive sign for  bosons  and negative for fermions) are the field
   {\em occupation numbers} in terms of the holomorphic variables $\widetilde{F}^{\pm}_\textbf{-l}$ defined by
    decomposition over
momenta \cite{JINR-1,grib80,ps1,Blaschke_04}:
 \begin{equation}\label{hh-1nt}
\widetilde{F}(\eta,\textbf{x})=\frac{1}{V_0}\sum_{\textbf{l},\textbf{l}^2\not
=0}c_F(a,\omega_{F,\textbf{l}})\frac{e^{i\textbf{\textbf{k}\textbf{x}}}}{\sqrt{2\omega_{F,\textbf{l}}}}
\left[\widetilde{F}^+_\textbf{l}(\eta)+\widetilde{F}^-_{-\textbf{l}}(\eta)\right],~~~~~~~
\textbf{k}=\frac{2\pi}{V^{1/3}_0}\textbf{l};
\end{equation}
 here
 $c_F(a,\omega_{F,\textbf{l}})$ is the normalization factor
determined by the Hamiltonian approach \cite{Barbashov_06}.
     Equations of these fields
   \bea\label{rel-2}
  \frac{\delta \widetilde{S}_{\rm U}[a|\widetilde{F}]}{\delta \widetilde{F}}=0
   \Big |^{\,d\eta=N(x^0)dx^0}_{\,a'^2=\widetilde{\rho}_{\rm U} (a)
   }
   \eea
  describe  a trajectory of a Relativistic Universe (RU)
     in its field space of events  $[a(\eta)|\widetilde{F}]$ \cite{WDW}
    as an analogy of the trajectory of a relativistic particle (\ref{rel-1}).
  The Hilbert foundations of relativistic physics \cite{H-1915}
  guarantee that the complete description of the  RU evolution can be fulfilled  by three ``times'':

 i) a relative``time-variable'' $a$ in the space of events $[a(\eta)|\widetilde{F}]$ as an object of
the frame  transformations,

 ii) an absolute ``time-interval'' $d\eta$ at the RU trajectory in
space of events, and

  iii) an unmeasurable ``time-coordinate'' $x^0$ in the Riemannian manifold as an object of
reparametrizations $x^0\to \widetilde{x}^0=\widetilde{x}^0(x^0)$.

  Really, the RU trajectory (\ref{rel-2}) is completed by its time-like component of
  $a'=\sqrt{\widetilde{\rho}_{\rm U}(a)}$ 
  that determines the coordinate-distance -- redshift relation $r(z)$ at the light-cone interval $\widetilde{ds}^2=d\eta^2-dr^2=0$
   \be\label{1-3}
 r(z)= \eta_0-\eta= \int\limits_{a=(z+1)^{-1}}^{a_0=1}\frac{da}{\sqrt{\widetilde{\rho}_{\rm U}(a)}}~.
 \ee
   This relation is  used as the basis of the cosmological
  studies of modern astrophysical data \cite{Giovannini,lif,MFB,Narlikar}.

 The best way to formulate the Relativistic Universe model in terms of reparametrization-invariant observables and their initial data is to use the Dirac Hamiltonian approach to the action (\ref{eta-1}) in the flat space-time approximation given by Eq. (\ref{cti-2}) \cite{Barbashov_06,dir}
 \bea\label{4-s}
 S \!= \!\! \int\limits_{}^{}\! d^4x
 \! \sum\limits_{F}^{} P_{\widetilde{F}}\partial_0\widetilde{F} 
+\int\!\!
 \left\{\!P_{\langle Q\rangle}d\langle Q\rangle\!+\!P_{\langle \phi \rangle} d\langle \phi \rangle
\! -\!P_{\log a}d\log a\!+\!\left[P_{\log  a}^2-\textsf{E}_{\rm U}^2(a)\right]
\frac{N_0(x^0)}{4V_0a^2}dx^0\right\},
 \eea
where $P_{\widetilde{F}}$,  $P_{\log a}=2V_0a a'$,
$P_{\langle \phi \rangle}=2a^2V_0\langle\phi\rangle'$,
and $P_{\langle Q \rangle}=2a^2V_0\langle Q\rangle'$
are the canonical conjugate momenta. In this case
\be\label{scm-4e}
  \textsf{E}_{\rm U}^2(a)\!\equiv\! P_{\langle \phi \rangle}^2\!+\!P_{\langle Q\rangle}^2\!+\!4V_0^2
 a^6\left[\textsf{V}_{\rm Higgs}(\langle\phi\rangle)+\textsf{V}_U(\langle Q\rangle)\right]+
 4V_0a^2 \textsf{H}_{\rm QFT}(\widetilde{F})
 \ee
can be  considered as a square of the Universe energy because $\log a$ is treated as
a Universe evolution parameter 
and  the global lapse function $N_0$ becomes the Lagrange multiplier. Variation of the latter leads to the energy constraint (\ref{free-c1a}) in its Hamiltonian form
  \be\label{m-0}
  P_{\log  a}^2-\textsf{E}_{\rm U}^2(a)=0.
  \ee

Quantities $\langle\phi\rangle$ and $\langle Q\rangle$ are the solutions of the equations
of motion following from the  emerging cosmological GR\&SM action (\ref{4-s}) with the initial
 data   for  the zeroth harmonics  (\ref {z-s1-1}) at the  instance
$\eta=0$:
\bea\label{in-d1}
\begin{array}{ll}
a(\eta=0)=a_I, \qquad \qquad
& P_{\log a_I}=\textsf{E}_{\rm U}(a_I),
\\\label{in-d2}
\langle\phi\rangle(\eta=0)=\dfrac{M_W}{g_W}\, ,
& P_{\langle \phi\rangle_I}=0,\\\label{in-d3}
\langle Q\rangle(\eta=0)=Q_I,
& P_{\langle Q\rangle_I}=2V_0H_0\sqrt{\Omega_{\rm rigid}}.
\end{array}
\eea
 These initial data as a natural source
  of  symmetry breaking in the Higgs doublet
 allow us to impose the zero potential conditions on the scalar zero modes
 $\textsf{V}_{\rm Higgs}(\langle\phi\rangle)\equiv 0$,
 $\textsf{V}_{\rm U}(\langle Q\rangle)\equiv0$~\cite{AGP_07}.

 However,  as it was revealed in \cite{Barbashov_06}, the  Wigner unitary irreducible representations of the Poincar\'e group, and
 the quantum  theory of a Relativistic Universe are not compatible with the
 accepted Lifshitz-Bardeen cosmological perturbation theory, where reparametrization-invariant
 conformal time is considered as an object of general coordinate transformations (frame transformations
 are not separated from the gauge ones) and the Hamiltonian approach is failed due to double counting
 of the scalar metric component. Moreover, the $\Lambda$CDM approach can be criticized because
it uses scalar metric components with a negative energy contribution
which have to be excluded, according to the Poincar\'e group classification of
 physical states of  metric components~\cite{Wigner-39}.

 The conformal variables $\widetilde{F},\widetilde{ds}$  reveals the fact that the scalar field  inflation scenario considered  in terms of  standard  variables  ${F},{ds}=a\widetilde{ds}$
    can be fulfilled
 only in the class of constant scalar fields $P_{\phi}\equiv 0$:  
 $$
S=-\int d\eta d^3x \,{a^4}\textsf{V}(\phi)
\Bigg |_{\phi=\mbox{\rm constant}}=-\int d\eta d^3x\,{a^4}\Lambda,
$$
where the scalar field potential produces the energy density $\rho_{\rm const.}={a^4}\Lambda \equiv H_0^2{a^4}{\Omega_{\Lambda}}$.
 Since the present-day value of the $\Lambda$-term is not equal to its initial data
 $\Lambda_I\not =\Lambda_0$, one needs the kinetic term too. However,  both
 the scalar field kinetic term and the  potential one
\be\label{md-1}
S=\int d\eta d^3x\left[{a^2}(\phi')^2-\lambda {a^4} \phi^4\right]\Big |^{P_c=P_\phi/{a}}_{\phi_c=\phi{a}}=\int d\eta d^3x\left[P_c \phi'-(\log {a})'P_c \phi_c -\left(\frac{P^2_c}{4}+\lambda \phi_c^4\right)\right],
\ee
 where $P_\phi=2a^2\phi'$, can lead to  integral of motion
 that corresponds to the  radiation  dominant conformal density~ ${\rho}={H_0^2}
 \Omega_{\rm radiation}$, instead of a sum of rigid state and inflation
 expected in  \cite{linde,mukh}. In particular, this formula
 signals   that dynamic scalar field inflation announced in \cite{linde,mukh} cannot exit.
In  this example (\ref{md-1}), the integral of motion shows us that the phenomenon of  dynamic scalar field inflation is an artefact of the choice of nonadequate  variables.
However,  formula (\ref{md-1}) contains the term $(\log {a})'P_c \phi_c$  as a source of an intensive creation of  scalar particles in the Early Universe revealed in \cite{Blaschke_04} for the longitudinal components of vector bosons. Thus, instead of dynamic scalar field inflation  \cite{linde,mukh}
we obtain effect of an intensive creation of  scalar particles in the Early Universe.

 In the following, we show that there is a possibility to solve the problems of the Wigner relativistic
 classification of physical states in modern Cosmology \cite{Wigner-39} (rejecting $\Lambda CDM$ model) and the absence of
 dynamic scalar field inflation announced in \cite{linde,mukh} using the Penrose-Chernikov-Tagirov dilaton gravity \cite{JINR-1} in order to
   describe of SN data and the CMB power spectrum.

\section{SN-Data in the Penrose-Chernikov-Tagirov Dilaton Gravity}

Recent data on the luminosity-redshift relation obtained by
the Supernova Cosmology project  \cite{snov,Riess_04} point out an
accelerated expansion of the universe within the standard
Friedman-Robertson-Walker (FRW) cosmological model. The accelerated expansion
can be achieved  by the $\Lambda$ term.
An alternative description of the
new cosmological supernova data without a $\Lambda$- term
was fulfilled in \cite{Behnke_02,Behnke_04,zakhy} using
 Weyl's geometry of similarity~\cite{we}.

As it was shown by Weyl~\cite{we} already in 1918, conformal
- invariant theories correspond to the relative standard of
measurement of a conformal - invariant ratio of two intervals
given in the geometry of similarity as a manifold of Riemannian geometries
connected by conformal transformations
 characterized by a measure of changing the length of
a vector on its parallel transport. The original
Weyl theory~\cite{we} was based on  the measure of changing the length of
a vector on its
 parallel transport as a vector field leading
to  the physical ambiguity of the arrow of time
pointed out by Einstein in his comment to Weyl's
paper~\cite{we}. However, in Dirac's paper \cite{Dir-73} and other~\cite{plb1,plb,kl,Tkach-04}
it was found that
the geometry of similarity can be realized in the case when the vector field
is replaced by the gradient of a scalar field -- dilaton $D$.
In this case, we obtain   the dilaton gravitational theory, i.e. Dilaton Gravity (DG).
The DG action $S_{\rm DG}$
coincides with  the Penrose-Chernikov-Tagirov action \cite{pct,JINR-1} with the opposite sign,
\be \label{dg-1}
S_{\rm DG}(D,\widetilde{g})=-S_{\rm PChT}(e^D,\widetilde{g})=-\int d^4x\left[\sqrt{-\widetilde{g}}\dfrac{e^{2D} R(\widetilde{g})}{6}-e^D\partial_\mu \left(\sqrt{-\widetilde{g}}\widetilde{g}^{\mu \nu}\partial_\nu e^D\right)\right].
\ee
This theory emerges from the Einstein one $S_{\rm GR}=-\int d^4x\sqrt{-{g}}R(g)/6$ after a scale transformation $S_{\rm DG}=S_{\rm GR}(g=e^{2D}\widetilde{g})$. However, the main difference of
DG from Einstein's
GR is the DG measurable interval $\widetilde{ds}^2={ds_1}^2/{ds_2}^2$ which is free from any
scale including the scale cosmological factor. In this case, the role of the scale cosmological factor is played by
the dilaton zero mode $\log a=\langle D\rangle$, so that in the cosmological
approximation the model given by the action (\ref{eta-1}) appears, where
conformal variables and coordinates  $\widetilde{F},\widetilde{ds}$ are identified
with measurable ones, and any dimensional parameter can be
introduced by the initial data. 

  The identification of measurable quantities with the conformal ones $\widetilde{F},\widetilde{ds}$
changes the numerical analysis of supernovae type Ia data \cite{snov,Riess_04}, because
instead of the Standard Cosmology (SC) distance -- redshift relation $R_{\rm SC}(z)=a(z)\int_{1}^{a(z)} d\overline{a} \rho_{\rm SC}^{-1/2}(\overline{a})$, $a(z)=1/(1+z)$ one uses
the Conformal Cosmology (CC) distance -- redshift relation
$R_{\rm CC}(z)=\int_{1}^{a(z)} d\overline{a} \rho_{\rm CC}^{-1/2}(\overline{a})$.

The   analysis in terms of the conformal {\em measurable} quantities~\cite{Behnke_02,Behnke_04,zakhy}
shows, in contrast to the $\Lambda$CDM model,
the dominance of the rigid state  ($\rho_{\rm CC}(a)=H_0^2\Omega_{\rm rigid}/a^2$,
$\sqrt{\Omega_{\rm rigid}}\sim 1$, $a(\eta)=\sqrt{1+2H_0(\eta-\eta_0)}$)
in all  epochs of the Universe evolution including the
chemical evolution, recombination,  and SN data. In the case of the  rigid state the conformal horizon
\be\label{1-3-w}
 \widetilde{d}(a)= 2 \int\limits_{0}^{a(z)=(z+1)^{-1}}\frac{d\overline{a}}{\sqrt{\widetilde{\rho}_{\rm CC}(\overline{a})}} =a^2H^{-1}_0
 \ee
coincides with the
inverse conformal Hubble
parameter
$\widetilde{d}_{\rm rigid}(a)=a^{2}H_0^{-1}=\widetilde{H}_{\rm rigid}^{-1}(a)$
for any value of the cosmological scale factor. This Conformal Cosmology \cite{Behnke_02,Behnke_04,zakhy}
is not excluded by modern observational data \cite{Riess:2001gk,Tegmark:2001zc}.

\section{Cosmological particle creation and the CMB power spectrum
}
 The main consequence of relativistic principles is
 the cosmological creation
 of primordial particles from vacuum \cite{JINR-1,JINR-2,Blaschke_04,Barbashov_06}
 as the origin of the Universe
 and its matter treated as ``Big Bang''.

The sources of vacuum creation of primordial particles  are
 in the linear differential form in  action (\ref{4-s})
  \bea\label{chh-2P}
\! \int d^3x \!\!\sum_{F} P_{\widetilde{F}}\partial_0\widetilde{F}\!=\!\frac{i}{2}\!\sum_{F,\textbf{l}}
  \left(\!\widetilde{F}^+_{-\textbf{l}}
  \partial_0\widetilde{F}^-_{\textbf{l}}\!-\!\widetilde{F}^-_{\textbf{l}}\partial_0\widetilde{F}^+_{-\textbf{l}}\right)
 \!+\!
  \frac{i}{2}\!\sum_{F,\textbf{l}}\!\!
  \left(\widetilde{F}^+_{-\textbf{l}}
  \widetilde{F}^+_{\textbf{l}}\!-\!\widetilde{F}^-_{\textbf{l}}\widetilde{F}^-_{-\textbf{l}}\right)
  \partial_0\widetilde{\triangle}_F.
 \eea
 In fact, the transformation~(\ref{hh-1nt}) to measurable quantities
 of the occupation number and one-particle energy leads to the second term on the right hand side of the above equation with  the following coefficients:
  $\widetilde{\triangle}_{F=v^{T},f}=\log\sqrt{\omega_F}$,
      $\widetilde{\triangle}_{F=v^{||}}=\log {a}/{\sqrt{\omega_F}}$,
  $\widetilde{\triangle}_{F=h,q}=\log {a}{\sqrt{\omega_F}}$,~$\widetilde{\triangle}_{F=Q,h^{TT}}=\log a$;
 here
 $v=v^{||}+v^{T}$ are conformal fields of $W$ and $Z$ vector bosons, $f$ are fermions, $h^{TT}$ is graviton, $h$ is a massive
 scalar (Higgs) particle (see  the massive vector theory in detail in \cite{Blaschke_04,hpp}).

 Due to the source terms  in the form (\ref{chh-2P})
 the field equations \be\label{fe-1}
   \partial_\eta \widetilde{F}^{\pm}(\textbf{k},\eta)= \pm i\omega_F\widetilde{F}^{\pm}(\textbf{k},\eta)
    +\partial_\eta \triangle_{F}(\eta)
   \widetilde{F}^{\mp}(\textbf{k},\eta)+i [\textsf{H}_{\rm int}, \widetilde{F}^{\pm}(\textbf{k},\eta)]
   \ee
 are not diagonal. In order to obtain integrals of motion in the approximation  $\textsf{H}_{\rm int}\sim 0$, these field equations are
 diagonalized by the  Bogoliubov transformation of the operator of particle
 $\widetilde{F}^+_{\textbf{l}}=\alpha b^+_{F,\textbf{l}}+\beta^* b^-_{F,\textbf{l}}$,
 so that the free equations of motion of the  Bogoliubov quasiparticle become diagonal
 $\partial_\eta b^{\pm}_{F,\textbf{l}}=\pm i\omega_b b^{\pm}_{F,\textbf{l}}$, where
  $\omega_b$ is the quasiparticle energy \cite{grib80,ps1,Blaschke_04}.
The stable vacuum is defined by $b^-_{F,\textbf{l}}|0>=0$,
where $b^-_{F,\textbf{l}}$ is the operator of annihilation of a quasiparticle.

One can assume \cite{ps1,Blaschke_04} that at the initial instance
$\eta=0$
 there were no any particle-like
excitations,  $ <0|\hat n_{\widetilde{F}}|0>(\eta=0)=0$, and hence the
temperature was equal to zero.
So the matter content of the Universe could appear as the final decay product
of the primordial vector bosons and the Higgs one created  from vacuum,
in accord with the Bogoliubov vacuum expectation value
\bea \label{free-1}
a'^2=\frac{H_0^2\Omega_{\rm rigid}}{a^2}+
\sum\limits_{F=h,Q,f,v}^{}\int \frac{d^3k}{(2\pi)^3}
\omega_{F,k}|\beta_{F,k}|^2B({\bf k},\widetilde{T}_{\rm F})
,
 \eea
where $\beta_{F,k}$ are the Bogoliubov coefficients  and
$ 
B({\bf k},\widetilde{T}_{\rm F})=\left\{
\exp\left[\dfrac{\sqrt{{\bf k}^2+m^2_{F0}a^2(z)}- m_{F0}a(z)}{ k_{\rm B}\widetilde{T}_{\rm F}}\right]
-1\right\}^{-1}
$ 
is the  distribution function arising due to interactions in (\ref{fe-1}) \cite{Smolyansky_02}.
This function coincides with the Boltzmann-Chernikov one \cite{CH-63,Muller-2007}, when the exponent is greater than unit.

The boson temperature $\widetilde{T}_{\rm F}\sim T_0$ can be estimated
 in the standard way from  the  relation  between   the free length
 $r_{\rm F}=[n(\widetilde{T}_{\rm F})\sigma_{\rm F~ scat}]^{-1}$, number density  $n(\widetilde{T}_{\rm F})$, and cross section $\sigma_{\rm F~ scat}$.
If  the free length is identified with   horizon
(\ref{1-3-w}), $r_{\rm F}\simeq \widetilde{d}(a)$, we have
  the collision integral kinetic equation
    \be\label{ke-1}
  \widetilde{n}(\widetilde{T}_{\rm F})\simeq[\widetilde{\sigma}_{\rm F~ scat}(a)\widetilde{d}(a)]^{-1}.
  \ee
 accepted in the present-day cosmological models \cite{Blaschke_04,ber-85}.

 In \cite{Blaschke_04}, the set of arguments was given in favor of that this formula (\ref{ke-1}), initial data, and SM hep-data give us all cosmological parameters, if   one supposes that the   rigid state $\rho_{\rm rigid}=p$ dominates for all epochs including the beginning of the Universe
 when  primordial vector bosons and Higgs particles were created from vacuum and
  their wavelengths  coincided with the horizon length, in particular
  \be\label{c-2a}  \widetilde{M}^{-1}_{Z\rm I} =[a_{Z\rm I} M_{Z}]^{-1}\simeq
\widetilde{H}_{Z\rm I}^{-1}=a^2_{Z\rm I} (H_{0})^{-1}~ \to~
 a_{Z\rm I}=(H_0/M_{Z0})^{1/3}=2.68 \cdot 10^{-15}.\ee
   One can suppose that the
collision integral kinetic equation (\ref{ke-1})
is valid for estimation of the present-day value
of the CMB temperature $\widetilde{T}_{\rm F}\sim T_0$, if CMB is considered  as the final decay product of the primordial  bosons created from vacuum.

 In order to describe the CMB anisotropy, the
collision integral kinetic equation (\ref{ke-1}) can be generalized
  to the anisotropic
  decays   $T_0\to T_0+\triangle T$, $\sigma_{\rm F~ scat}(a)\to \sigma_{\rm F~ scat}(a)+\sigma_{2\gamma}$, so that we obtain
  a formula
 \be\label{c-5}\Big|\frac{\triangle T}{T_0}\Big|\simeq
 \frac{1}{3}\Big|\frac{\sigma_{2\gamma}}{\sigma_{\rm F~ scat}(a)}\Big|\sim \alpha_{\rm QED}^2
 \sim 10^{-5}.
 \ee
that allows us to establish processes that form the CMB anisotropy
using   the observational value of the CMB anisotropy
 $\sim 10^{-5}$  \cite{WMAP01,hu01,WMAP08}.
This value
 testifies  to the dominance of the two photon processes. Therefore,
the CMB anisotropy revealed in \cite{WMAP01,hu01,WMAP08} in the region of three
peaks $\ell_1\simeq 220$,  $\ell_2\simeq 546$, and $\ell_3\simeq800$
can reflect rather parameters of the primordial bosons and their
decay processes, in particular
 $h\to \gamma\gamma$, $W^+W^-\to
\gamma\gamma$, and $ZZ \to \gamma\gamma$, than the ones of
 matter at the time of recombination, as it is accepted in  the $\Lambda$CDM model where
the scalar metric components are used  as dynamical
variables  \cite{MFB} which are not compatible with the Poincar\'e group classification of
physical states \cite{Barbashov_06,B_06,Glinka_08}. The spectrum
of the Higgs and vector boson masses $m_h, M_Z, M_W$ can be obtained from
the CMB power spectrum using the Gamov
identification of the energy of the processes with the product of their redshift z-factors $(z_P+1)$
and the present-day CMB temperature $T_0=2.725$ K = $2.35\cdot 10^{-13}$ GeV
\be\label{cr-7a}
M_{P0}=T_0(z_P+1).
\ee
The z-factors $(z_P+1)$ of the present day processes energy can be expressed through
 the redshift
\be\label{cr-7-2}
(Z_{Pd}+1)=(z_P+1)a_{\rm L}
\ee
of their primordial values $\widetilde{M}_{PI}=M_{P0}\,a_{PI}$,  where $a_{PI}$ is
defined by Eq. (\ref{c-2a}) and
 $a_{\rm L}$ is
 the universal factor  for all processes
 characterizing the CMB spectrum at the time
of its establishment.

 The  initial data reference redshift $(Z_{Pd}+1)$
is defined by
values of multipole momenta at the CMB anisotropic peaks that can be obtained  using the
accepted formula  \cite{hu01}
\be\label{cr-6a}
\ell_{Pd}= \widetilde{d}(a_{Pd}) \widetilde{M}_P(a_{Pd})={d}(a_{Pd}) M_{P0}=\frac{a_{Pd}^3}{a_{PI}^3} =(Z_{Pd}+1)^3,
\ee
where   $\widetilde{d}(a_{Pd})=a_{Pd}^2H_0^{-1}$ is the conformal
horizon  (\ref{1-3-w}) at the instances of the two photon processes $(P)$
marked by the cosmological scale factors $a_{Pd}$ and $a^3_{PI}=H_0/M_{P0}$ in accord with Eq. (\ref{c-2a}).

Using formulae (\ref{cr-7a}) -- (\ref{cr-6a})
 one can obtain
the final  formula that expresses  the boson spectrum through the power spectrum of the CMB multipole momenta
\be\label{cr-7}
M_{P0}=T_0(z_P+1)=T_0a^{-1}_{\rm L}(Z_{Pd}+1)=T_{\rm L}(Z_{Pd}+1)=T_{\rm L}\ell_{Pd}^{1/3},\ee
where $T_{L}\simeq 9.8 \mbox{\rm GeV}$ is defined by  the boson masses:
$\ell^{1/3}_{Wd}=\dfrac{M_{W0}}{T_L}$, $\ell^{1/3}_{Zd}=\dfrac{M_{Z0}}{T_L}$,
so that
\be
\frac{M_Z}{M_W} = 1.134 \approx \left(\frac{\ell_3}{\ell_2}\right)^{1/3}= \left(\frac{800}{546}\right)^{1/3}= 1.136.
\ee

Formula  (\ref{cr-7}) predicts the Higgs mass as
\be
m_h=2M_W\left(\frac{\ell_1}{\ell_2}\right)^{1/3}=2M_W\left(\frac{220}{546}\right)^{1/3}\simeq 118 \mbox{\rm GeV}.
\ee
This value of the Higgs boson mass is close to the present fit of the LEP
experimental data supporting rather low values of  the experimental
limit $m_h>114.4$~GeV \cite{abpz-08}.

\section{The Relativistic Universe Scenario Problems}

In context of relativistic physics the topical problems of modern Cosmology
 are the following:

 1. the acoustic
 explanation of the CMB power spectrum by the dynamical scalar metric component is not compatible with
 the relativistic classification of physical states, because the latter do not contain this component.

 2. the absence of the dynamical scalar field inflation (as it was shown in Eq. (\ref{md-1}),

 3. the lost of physical meaning of concepts of the ``particle mass'' $M_P$ and ``temperature'' $T(a)$
 at the limit of small horizon  before the Planck epoch, when $H(a)\geq M_P, T(a)$.

 We discussed possibilities to solve these problems by
 the Chernikov-Parker cosmological vacuum creation of vector bosons and the Higgs particle
with  Chernikov kinetic theory of relativistic gas in the Penrose-Chernikov-Tagirov Dilaton Gravity.

However, in order to realize these possibilities, one should consider the following
problems.

{\bf 1. The Hamiltonian approach to the conformal-invariant unified theory}.

Like the Copernican assertion that {\it  we can measure
 only a difference of coordinates}  pointed out the Galilei
  pathway to the Newton mechanics,
 the Weyl assertion that {\it  we can measure
 only a ratio of two intervals} can point out a pathway
   to one of the conformal-invariant unified theories.

 The RU model is based on the Penrose-Chernikov-Tagirov conformal-invariant unified theory \cite{pct,JINR-1}
 \be \label{ru-1}S_{\rm RU}[D,\widetilde{F}^{(n)}]=S_{\rm U}[F^{(n)}=e^{nD}\widetilde{F}^{(n)}]\ee
where $S_{\rm U}[F^{(n)}]$ coincides with the sum of GR and SM (\ref{1-1})
 with the initial data given in the  CMB frame of reference, where the physical variables and fields are chosen.
 The CMB frame is given in   the space-time
 with the finite volume  $\int d^3x=V_0$ (compatible with the dark sky at night \cite{DarkS}), the geometric interval
 \bea \label{dg-2}\widetilde{ds}^2&=&\widetilde{\omega}^2_{(0)}-\widetilde{\omega}^2_{(b)},\\ \label{dg-3}
 \widetilde{\omega}_{(0)}&=&e^{2{D}}{N}_ddx^0=e^{2{D}}{\cal N}{}d\tau,\\\label{dg-4}
 \widetilde{\omega}_{(b)}&=&{\bf e}_{(b)j}(dx^j+N^j dx^0)={\omega}^{(3)}_{(b)}+{\cal N}_{(b)} d\tau,
 \eea
  the unit spatial metric determinant $|{\bf e}_{(b)i}|=1$, and
   the particle masses $m(D)=e^{{D}}m_0$. In accord with Einstein's cosmological principle (\ref{z-s1-1}) \cite{Einstein_17} zeroth harmonics of  all scalar fields including the dilaton one $\langle D\rangle=V_0^{-1}\int d^3x D=\log a$ are 
       separated  $D=\langle D\rangle+\overline{D}$.

  Just this separation allows us to identify $\langle D\rangle$ with the logarithm of the cosmological scale factor $a$ (considered in Cosmology as the evolution parameter
  in the field space of events) and introduce the Misner diffeo-invariant time-interval defined as \cite{M}
 \be\label{1-4dg}
 d\tau=\frac{d\eta}{a^2(\eta)}=dx^0\langle N_d^{-1}\rangle^{-1}\equiv dx^0N_0,~~~~~~~~~\frac{df}{d\tau}=\partial_\tau f = \left\langle\frac{df}{N_ddx^0}\right\rangle=V_0^{-1}\int d^3x\frac{df}{N_ddx^0}.
 \ee

     The action  (\ref{ru-1}) after the separation of cosmological scale factor $a=e^{\langle D\rangle}$ coincides with the corresponding action of the type of (\ref{eta-1}) obtained from (\ref{1-1}).
    In particular, the Dilaton Gravity  (\ref{ru-1})  with a set of free scalar fields $D,\phi,Q$ takes the form
  \bea\label{1-1dg}
 S^{\rm S}_{\rm RU}[\widetilde{g}|D,\phi,Q]&=&\int\limits_{ }^{ }d^4x\sqrt{-g}\left[-\frac{1}{6}R(g)
 +\partial_\mu \phi\partial^\mu \phi +\partial_\mu Q\partial^\mu Q \right]\bigg |_{g=e^{2D}\widetilde{g}}\\\label{1-2dg}
 &\equiv&
 \widetilde{S}^{\rm S}_{\rm RU}
 +\underbrace{V_0\int\limits_{\tau=0}^{\tau_0} d\tau [-\langle \partial_\tau D\rangle^2+\langle \partial_\tau\phi \rangle^2+\langle \partial_\tau Q \rangle^2]}_{zeroth-mode~contribution}=\widetilde{S}^{\rm S}_{\rm RU}+
 S_{\rm zero}^{\rm S},
 \eea
  where
  \bea\label{1-3dg}
 \widetilde{S}^{\rm S}_{\rm RU}&=&\int\limits_{\tau=0}^{\tau_0} dx^0 \int d^3x N_{\rm d}
 \left[-{v^2_{\overline{D}}}+\frac{v^2_{(a)(b)}}{6}-e^{4D}\frac{ R^{(3)}({\bf e})+8e^{-D/2}\triangle e^{D/2}}{6}\right]\\
 &+&\int\limits_{\tau=0}^{\tau_0} dx^0 \int d^3x N_{\rm d} \left[v_{\overline{\phi}}^2 -\left(\partial_{(a)}\overline{\phi}\right)^2+v_{\overline{Q}}^2 -\left(\partial_{(a)}\overline{Q}\right)^2\right]
 \eea
 is the action of the local field variables,
  \bea\label{proi11}
 v_{\overline{D}}&=&\frac{1}{{N_d}}\left[(\partial_0-N^l\partial_l)\overline{D}
 -\frac13\partial_lN^l
\right],\\\label{proi12}
 v_{(ab)}&=&\frac{1}{2{N_d}}\left({\bf e}_{(a)i}v^i_{(b)}+{\bf e}_{(b)i}v^i_{(a)}\right),\\\label{proizvod1}
 v_{(a)i}&=&
 \frac{1}{{N_d}}\left[(\partial_0-N^l\partial_l){\bf e}_{(a)i}
+ \frac13 {\bf
 e}_{(a)i}\partial_lN^l-{\bf e}_{(a)l}\partial_iN^l\right],\\\label{sc-1}
 v_{\overline{\phi}}&=&\frac{1}{{N_d}}\left[\partial_0\overline{\phi}-{N}^k\partial_k\overline{\phi}\right]
 \\\label{sc-2}
 v_{\overline{Q}}&=&\frac{1}{{N_d}}\left[\partial_0\overline{Q}-{N}^k\partial_k\overline{Q}\right]
 \eea
 are velocities of the metric components and scalar fields,
   $\triangle\psi=\partial_i({\bf e}^i_{(a)}{\bf
 e}^j_{(a)}\partial_j\psi)$ is the covariant Laplace operator,
 ${}^{(3)}R({\bf{e}})$ is a three-dimensional curvature
 expressed in terms of triads
   ${\bf e}_{(a)i}$:
\be \label{1-17}
 {}^{(3)}R({\bf e})=-2\partial^{\phantom{f}}_{i}
 [{\bf e}_{(b)}^{i}\sigma_{{(c)|(b)(c)}}]-
 \sigma_{(c)|(b)(c)}\sigma_{(a)|(b)(a)}+
 \sigma_{(c)|(d)(f)}^{\phantom{(f)}}\sigma^{\phantom{(f)}}_{(f)|(d)(c)}.
 \ee
 Here
 \be\label{1-18} \sigma_{(a)|(b)(c)}=
 {\bf e}_{(c)}^{j}
 \nabla_{i}{\bf e}_{(a) k}{\bf e}_{(b)}^{\phantom{r}k}=
 \frac{1}{2}{\bf e}_{(a)j}\left[\partial_{(b)}{\bf e}^j_{(c)}
 -\partial_{(c)}{\bf e}^j_{(b)}\right]
  \ee
  are the coefficients of the spin-connection (see \cite{B_06}
  ),
  $\nabla_{i}{\bf e}_{(a) j}=\partial_{i}{\bf e}_{(a)j}
  -\Gamma^k_{ij}{\bf e}_{(a) k}$~are covariant derivatives, and
  $\Gamma^k_{ij}=\dfrac{1}{2}{\bf e}^k_{(b)}(\partial_i{\bf e}_{(b)j}
  +\partial_j{\bf e}_{(b)i})$.

  The Wigner-Poincar\'e classification  excludes physical states
  with negative contribution of the kinetic energy of the type of  local dilaton
  $-v^2_{\overline{D}}$ in action (\ref{1-1dg}). This  means that one can impose the additional constraint
  of zero canonical momenta of the local dilaton field
  \be\label{1-5dg}
  p_{\overline{D}}=-2v_{\overline{D}}\simeq 0. 
  \ee
 The last constraint, in the Dirac approach to GR \cite{dir,Barbashov_06},
means  that the velocity of the spatial volume
   element is equal to zero,
    and it leads to the positive
   Hamiltonian density.

   This Dirac constraint (\ref{1-5dg}) is the consequence of
   the relativistic  classification   of physical states in the theory (\ref{ru-1}).
      This  classification
   is not compatible with both the $\Lambda CDM$ model,
   and the Inflationary Model \cite{linde,mukh}. In these models the CMB
    (treated as an object of relativistic  transformations) is described by the dynamical scalar component
   (forbidden by the relativistic  classification of physical states).

    The
   calculation of all canonical momenta including (\ref{1-5dg}) and
\bea\label{m-1}\widetilde{p}_{(ab)}&=&\frac{1}{2}({\bf e}^i_{(a)}\widetilde{p}_{(b)i}+
  {\bf e}^i_{(b)}\widetilde{p}_{(a)i})=\frac{1}{3}v_{(a b)},
 \\\label{m-2}
\widetilde{p}^i_{(b)}&=&\frac{1}{3}{\bf e}^i_{(a)} v_{(a b)},~~~  p_{\overline{\phi}}=2v_{\overline{\phi}}, ~~~p_{\overline{Q}}=2v_{\overline{Q}}, \\\label{m-3}
  P_{D}&=&2V_0 \partial_\tau \langle D\rangle,~~~
P_{\langle \phi \rangle}=2V_0\langle\partial_\tau\phi\rangle,~~~
P_{\langle Q \rangle}=2V_0\langle \partial_\tau Q\rangle
 \eea
 allows us to represent
  the action  (\ref{ru-1})
   in the Hamiltonian form \cite{B_06,Glinka_08}
 \bea\label{h-1c}
 S^S_{\rm DG}&=&\int dx^0\int d^3x\left[\widetilde{p}^i_{(b)}\partial_0{\bf e}^i_{(b)}-p_{\overline{D}}\partial_0\overline{D}+
 p_{\overline{\phi}}\partial_0\overline{\phi}+p_{\overline{Q}}\partial_0\overline{Q}
 -{N_d} T_{\rm d}
 - \textsf{C}\right]\\\label{h-1g}
 &+&\int\!\!
 \left\{\!P_{\langle Q\rangle}d\langle Q\rangle\!+\!P_{\langle \phi \rangle} d\langle \phi \rangle
\! -\!P_{\langle D\rangle}d\langle D\rangle\!+\!\left[\frac{P_{\langle D\rangle}^2-P_{\langle \phi \rangle}^2-P_{\langle Q\rangle}^2}{4V_0}\right]
{N_0(x^0)}dx^0\right\},
 \eea
 where
 \bea
\label{3L+1G-8}
 {T}_{\rm d}&=&\dfrac{4}{3}{e}^{7D/2}
 \triangle
{e}^{D/2}+
  \sum\limits_{I=0,2,3,4,6} {e}^{ID}{\cal T}_I
 \eea
 is the sum of local field energy  densities  ${\cal T}_I$ including the gravity density
  defined in  \cite{Barbashov_06,B_06,Glinka_08}, and
\be\label{h-2c}
 \textsf{C}=N_{(b)}
  {T}^0_{(b)} +\lambda_0{p_{\overline{D}}}+ \lambda_{(a)}\partial_k{\bf e}^k_{(a)}
 \ee
 is a sum of the Dirac constraints with corresponding Lagrangian multipliers, including
 three   first class constraints
 \bea\label{h-c1}
 T_{(0)(a)}&=&
  -p_{\overline{D}}\partial_{(a)}
 {\overline{D}}+\frac{1}{3}\partial_{(a)}
 (p_{\overline{D}}\overline{D}) +
 2p_{(b)(c)}\sigma_{(b)|(a)(c)}-\partial_{(b)}p_{(b)(a)}+
 p_{\overline{\phi}}\partial_{(b)}\overline{\phi}+p_{\overline{Q}}\partial_{(b)}\overline{Q},
  \eea
 and the fourth second class ones
  \bea\label{1-42}
 \partial_k{\bf e}^k_{(a)}&=&0,\\\label{1-43}
 p_{\overline{D}}&=&0 ~~~\to~~~
 \partial_\tau e^{3\overline{D}}=
 \partial_{(b)}\left(e^{3\overline{D}}{\cal N}_{(b)}\right),
 \eea
where ${\cal N}_{(b)}={N}_{(b)}/N_0$, $N_0$ is given by Eq. (\ref{1-4dg}). The latter (\ref{1-43}) is the analog of
  the Dirac condition
 of the minimal 3-dimensional hyper-surface \cite{dir} in GR that gives a positive value of the Hamiltonian density.

       We can see that the accepted Einstein Eqs.
      \bea\label{E-1}
      -{e}_{(\nu)\alpha}
\frac{\delta S^{\rm S}_{\rm RU}}
{\delta {e}_{(\mu)\alpha}}\equiv T_{(\nu)(\mu)}
 =0
      \eea
      coincide with the Einstein equations in infinite volume beside
 the energy constraint
      \bea\label{h-c1d}
 -\frac{\delta S^{\rm S}_{\rm RU}}{\delta {N}_{\rm d}}
 &=&\frac{P_{\langle D\rangle}^2-P_{\langle \phi \rangle}^2-P_{\langle Q\rangle}^2}{4{N}_{\rm d}V^2_0}-{N}_{\rm d} T_{\rm d}=0.\eea
 One can see that the Einstein equations (\ref{E-1}) play  a role of a set geometrical constraints, rather than the energy-momentum tensor components. In particular,
  the energy constraint (\ref{h-c1d}) shows us how the Einstein cosmological principle (\ref{z-s1-1})
  solves the problem of energy in finite space-time by
  averaging energy constraint (\ref{h-c1d}) over the volume $V_0$. This averaging
  defines the Universe energy   as one of the constraint-shell values of the evolution parameter
  momentum $P_{\langle D\rangle}$
 \bea\label{4d}
 \textsf{E}_{\rm U}^2&\equiv& P_{\langle D\rangle}^2= P_{\langle \phi \rangle}^2+P_{\langle Q\rangle}^2+
 4 \left[\int d^3x \sqrt{T_{\rm d}}\right]^2
 \eea
 in accord with the Wigner-Poincar\'e classification of relativistic physical states
  in the CMB frame. This definition  leads to the luminosity-distance -- redshift relation (\ref{1-3})
  in the
  case of the  rigid state dominance ($P_{\langle Q\rangle}\gg  \int d^3x \sqrt{T_{\rm d}}$)
  \be\label{1-3-w-d}
 \ell(z)=H_0\frac{r(a)}{a^2}= \frac{V_0}{a^{2}} \int\limits_{0}^{a(z)=(z+1)^{-1}}\frac{d\overline{a}^2}{\textsf{E}_{\rm U}} =z+\frac{z^2}{2}.
 \ee
 This luminosity-distance -- redshift relation does not contradict to
 the recent SN data
  \cite{Riess_04} in the framework of the Conformal Cosmology \cite{Behnke_02,Behnke_04,zakhy}.
 The substitution of the energy (\ref{4d}) into the energy constraint (\ref{h-c1d}) gives us
  the diffeo-invariant local lapse function
 \bea
\label{3d}
 {\cal N}&=&
 {\langle{T}_{\rm
 d}^{1/2}\rangle}{{T}_{\rm d}^{-1/2}}
\eea
  in the interval  (\ref{dg-4}), where
   the dilaton field $\overline{D}$ is defined by equations
 \bea
 \label{3L+1G-3}
 -\frac{\delta S}{\delta D}&=&0
 ~~~ \Rightarrow~~~
 \partial_\tau P_{\langle {D}\rangle}=\langle {T}_{D}\rangle,~~~{T}_{D}-\langle {T}_{D}\rangle=0,
 \eea
    \bea
\label{3L+1G-8d}
  {T}_{D}&=&\dfrac{2}{3}
   \left\{7{\cal N}{e}^{7D/2}
  \triangle {e}^{D/2}+{e}^{D/2} \triangle
\left[{\cal N}{e}^{7D/2}\right]\right\}+
  {\cal N}\sum\limits_{I=0,2,3,4,6}I {e}^{ID}{\cal T}_I.
 \eea
 In the  case of the scalar fields (massless or massive)
 with initial data (\ref{in-d1}) 
 the theory is reduced to the version of the Universe considered before in previous Section with
 cosmological creation of the   scalar fields (massless or massive ones if there are potentials),
 where the collision integrals and temperature
is completely determined by
  the Einstein equations, their initial data, and constraints (\ref{1-42}) --
 (\ref{3L+1G-8d}). 
 It was shown that the cosmological
 perturbation theory for ($e^D, {\cal N}, {\cal N}_{(b)}$) is the cosmological generalization of the Schwarzschild solution \cite{Barbashov_06}.

 In this case the Universe is a factory of  scalar particles created in the homogeneous background of their
 zeroth modes.
 The densities (\ref{3L+1G-8}) and (\ref{3L+1G-8d}) keep only  component $I=2$ of the radiation or massive one $I=3$. Their parameters can be approximately described  by the collision integral (\ref{ke-1}).
The problem is to consider this scenario   in the SM.

{\bf 2. The initial data Higgs effect in Cosmology}.

In  DG, the Higgs effect is provided by the initial data (\ref{in-d1}).
In particular, the $U(1)$  model  in the cosmological approximation $D\simeq \langle D\rangle=\log a$ is given by the action and Lagrangian
\bea\label{h-1}S_1&=&\int d\eta\int d^3x{\cal L}_1,\\
{\cal L}_1&=&-\frac{1}{4}[\partial_\mu A_\nu-\partial_\nu A_\mu]^2 +a^2|(\partial_\mu \phi -ieA_\mu)\phi|^2
+\overline{\widetilde{\psi}} (i \hat \partial +e \hat A +a|\phi|)\widetilde{\psi}-a^4\textsf{V}_{eff} (|\phi|).
\eea
After transition to physical variables  $\phi =e^{i\chi}|\phi|, \widetilde{\psi}=e^{-i\chi}f, A\mu=B_\mu-(1/e)\partial_\mu \chi$
this $U(1)$-Lagrangian takes the form
\be\label{h-2}
{\cal L}_1=-\frac{1}{4}[\partial_\mu B_\nu-\partial_\nu B_\mu]^2 +e^2a^2|\phi|^2(B_\mu)^2+a^2(\partial_\mu|\phi|)^2
+\overline{f} (i \hat \partial +e \hat B +a|\phi|)f-a^4\textsf{V}_{eff} (|\phi|),
\ee
where $a|\phi|=|\widetilde{\phi}|=a\langle \phi\rangle+\widetilde{h}/\sqrt{2}$ and $\int d^3x h(\eta,x)=0$.
The consistency conditions
 \bea
 \label{va-2}
 \textsf{V}_{eff} (\phi_{\rm I})=0,~~~
 \frac{d\textsf{V}_{eff}(\langle\,\phi\rangle)}{d\langle\,\phi\rangle}~
 \Bigg|_{\langle\,\phi\rangle=\phi_{\rm I}}=0.
 \eea
 should be imposed in order to keep free dynamics of the zero component $\langle\phi\rangle$
  as a  solution of the variational equation $[a^2\langle\phi'\rangle]'=0$ with the initial data
   $\langle\phi'\rangle (\eta=0)=0$ and $\langle\phi\rangle (\eta=0)=\phi_{\rm I}$ \cite{AGP_07}.

%

   The problem is to fulfill this analysis in  SM.

{\bf 3. Kinetic theory of the primordial particle vacuum creation}.

  To get a more accurate estimate of the Higgs mass and a better description of
the CMB power spectrum within the model under consideration, one has to
perform an involved analysis of the kinetic equation \cite{CH-63} for
nonstationary processes of primordial particle creation and
subsequent decays.

In particular, the lifetime $\eta_L$ of
 product bosons in the early Universe in dimensionless units
$\tau_L=\eta_L/\eta_I$, where $\eta_I=(2H_I)^{-1}$, can be estimated
by using the equation of state $a^2(\eta)=a_I^2(1+\tau_L)$ and the
$W$-boson lifetime within the Standard Model. Specifically, we have
\be \label{life} 1+\tau_L= \frac{2H_I\sin^2
\theta_{(W)}}{\alpha_{\rm QED} M_W(\eta_L)}= \frac{2\sin^2
\theta_{(W)}}{\alpha_{\rm QED}\sqrt{1+\tau_L}}, \ee
where $\theta_{(W)}$ is the Weinberg angle,  $\alpha_{\rm
QED}=1/137$ is the fine-structure constant, and $
M_{I}/ H_I\simeq 1$.

From the solution to Eq.~(\ref{life}), $ \tau_L+1=
\left({2\sin^2\theta_{(W)}}/{\alpha_{\rm
QED}}\right)^{2/3} \simeq {16}$ it
follows that the lifetime of product bosons
is an order of magnitude longer than the Universe relaxation time:
\be \label{lv} \tau_L ={\eta_L}/{\eta_I}\simeq
16-1=15. \ee

The problem is to obtain
the parameters of the diffusion reaction system
arising in this case from the Standard Model
computing the relevant cross sections and decay rates.

{\bf 4. Baryon-antibaryon asymmetry of matter in
the Universe}.

In SM, in each of the three generations of leptons
 (e,$\mu$,$\tau$) and color quarks, we have four fermion
 doublets -- in all there are $n_L=12$ of them. Each of 12 fermion
 doublets interacts with the triplet of non-Abelian fields
 $A^1=(W^{(-)}+W^{(+)})/\sqrt{2}$, $A^2=
 i(W^{(-)}-W^{(+)})/\sqrt{2}$, and $A^3=Z/\cos\theta_{(W)}$,
 the corresponding coupling constant being $g=e/\sin\theta_{(W)}$.
 It is well known that, because of a triangle anomaly, W- and
Z- boson interaction with lefthanded fermion doublets
 $\psi_L^{(i)}$, $i=1,2,...,n_L$, leads to
 nonconservation of the number of fermions of each
type ${(i)}$
\cite{bj,th,ufn},
 \bea \label{rub}
 \partial_\mu j^{(i)}_{L\mu}=\frac{1}{32\pi^2}
 {\rm Tr}\hat F_{\mu\nu}{}^*\!{\hat F_{\mu\nu}},
 \eea
 where $\hat
 F_{\mu\nu}=-iF^a_{\mu\nu}g_W\tau_a/2$ is the strength of the
 vector fields, $F^a_{\mu\nu}=
 \partial_\mu A_\nu^a-\partial_\nu
 A_\mu^a+g\epsilon^{abc}A_\mu^bA_\nu^c$.

 Taking the integral of the equality in (\ref{rub}) with respect to conformal time and
 the three-dimensional variable $x$, we can find a relation between
 the change
 \be \label{rub-1}
 \int\limits_{\eta_I}^{\eta_0} d\eta \int d^3x
 \partial_\mu j^{(i)}_{L\mu}=F^{(i)}(\eta_0)-F^{(i)}(\eta_I) =\Delta F^{(i)}\ee
  of the fermion number $ F^{(i)}=\int d^3x
 j_0^{(i)}$ and the Chern-Simons functional
 $ 
 F_{\mu\nu}{}^*\!{\hat F_{\mu\nu}},
 $ 
 so that after integration Eq.  (\ref{rub}) takes the form
  \be \label{rub2}
 \Delta F^{(i)}= N_{CS} \not = 0, ~~~i=1,2,...,n_L.
 \ee
 The equality in (\ref{rub2}) is considered as a selection rule --
that is, the fermion number changes identically for all fermion
types:
 $N_{CS}=\Delta L^e=\Delta L^\mu=\Delta L^\tau=\Delta B/3$;
 at the same time, the change in the baryon charge $B$ and the change
 in the lepton charge  $L=L^e+L^\mu+L^\tau$ are related to each other in
such a way that $B-L$ is conserved, while
 $B+L$ is not invariant. Upon taking the sum of the equalities in
 (\ref{rub2}) over all doublets, one can  obtain $\Delta (B+ L)=12
 N_{CS}$ ~\cite{ufn}.

 We can evaluate the expectation value of the Chern-Simons
 functional (\ref{rub2})  (in the lowest order of perturbation
 theory in the coupling constant) in the Bogoliubov vacuum
 $b|0>=0$. Specifically, we have
 \be
 N_{CS}=N_{\rm W}\equiv
 -\frac{1}{32\pi^2}\int_0^{\eta_{L}} d\eta \int {d^3 x} \;
 \langle 0|{\rm Tr}\hat F^{\rm W}_{\mu\nu}
 {}^*\!{\hat F^{\rm W}_{\mu\nu}}|0\rangle ,
 \ee
 where $\eta_{L}$  is the W-boson
 lifetime, and $N_{\rm W}$
 is the contribution
 of the primordial $W$  boson.
 The integral over the conformal spacetime bounded
 by three-dimensional hypersurfaces $\eta=0$ and $\eta =\eta_L$
  is given by
 $N_{\rm W} ={2}{\alpha_W}V_0
 \int_{0}^{{\eta_{L}}} d\eta \int\limits_{0 }^{\infty }dk
 |k|^3 R_{\rm W}(k,\eta)
 $,
 where
 ${\alpha_W}={{\alpha}_{\rm
 QED}}/{\sin^{2}\theta_{W}}$
 and
 $R_{\rm W}=\frac{i}{2}{}_b<0|b^+b^+-b^-b^-|0>_b=-\sinh(2r(\eta_L))\sin(2\theta(\eta_L))
 $
  is the Bogoliubov condensate \cite{Blaschke_04} that is
 specified by relevant solutions to the
 Bogoliubov equations. Upon a numerical calculation
 of this integral, we can estimate the expectation value of the
 Chern-Simons functional in the state of primordial bosons.

 At the vector-boson-lifetime values of
 $\eta_{L}= 15$,
 this yields the following result at 
 $n_{\gamma}={ 2,402  \times T^3 }/{\pi^2}$
\be
\frac{N_{CS}}{V_{0}}=
\frac{ N_W}{V_{0}}\\
 = 4\alpha_W 
  T^3
 \times 1.44
  =0.8~  n_{\gamma}.
\ee
where $n_{\gamma}$ is the number
 density of photons forming Cosmic Microwave Background radiation.
On this basis, the violation of the fermion-number density in
the cosmological model being considered can be estimated as
\cite{Blaschke_04,Behnke_02}
$
{\Delta F^{(i)}}/{V_{0}}={N_{CS}}/{V_{0}}
  =0.8  n_{\gamma}
$. 

 According to SM, there is the CKM-mixing
 that leads to ${\rm CP}$ nonconservation,
 so that the cosmological evolution and
 this  nonconservation freeze
  the
  fermion number at $\eta=\eta_L$.  This
 leads to the baryon-number density \cite{ufn,sufn}
$ 
  n_{\rm b}(\eta_L)=
  X_{\rm CP}{\Delta
  F^{(i)}}/{V_{0}}\simeq X_{\rm CP}n_{\gamma}(\eta_L)
$,  
  where the factor $X_{\rm CP}$ is determined by the superweak
 interaction of $d$ and $s$ quarks,
 which
 is responsible for CP violation experimentally observed in
 $K$-meson decays \cite{o}.

 From the ratio of the number of baryons to the number of photons,
 one can deduce an estimate of the superweak-interaction coupling
 constant: $X_{\rm CP}\sim 10^{-9}$.
  Thus, the evolution of the Universe, primary
 vector bosons, and the aforementioned superweak interaction~\cite{o}
  lead to baryon-antibaryon
 asymmetry of the Universe
 \be\label{data6a} \frac{n_{\rm b}(\eta_L)}{n_{\gamma}(\eta_L)}\simeq X_{\rm CP}= 10^{-9}.
  \ee


 Thus, the primordial bosons before
 their decays polarize the Dirac fermion vacuum and give the
 baryon asymmetry frozen by the CP -- violation
 so that for billion photons there is only one baryon.

The problem is to show that the Universe matter content
considered as the final decay product of primordial bosons is in agreement with
observational data~\cite{Blaschke_04}.

{\bf 5. Nucleosynthesis in the Early Universe}.

Calculation of the primordial helium abundance \cite{Wein_73,Behnke_04} takes into account
  weak interactions,
 the Boltzmann factor, (n/p) $ e^{\triangle m/T} \sim 1/6$, where $\triangle m$ is the neutron-proton
mass difference, which is the same for both SC and CC,
$\triangle m_{SC}/T_{SC}=\triangle m_{CC}/T_{CC}=(1+z)^{-1}m_{0}/T_{0}$,
and the square root dependence of the z-factor on the measurable time-interval.

In SC, where the measurable time-interval is identified with
the Friedmann time, this square root dependence of the z-factor is explained by the
radiation dominance, whereas in Conformal Cosmology, where the measurable time-interval is identified with
 the conformal time $\eta$, the square root dependence of the z-factor
is  explained by the dominant rigid state $(1+z)^{-1}= \sqrt{1+2H_0(\eta-\eta_0)}$.
Therefore, at first sight the Conformal Cosmology model with  dominant rigid state does not contradict to
the primordial helium abundance \cite{Wein_73,Behnke_04}.
This rigid state describes both the cosmological creation of primordial particles
with the CMB power spectrum (as we have seen above) and the Luminosity-distance -- redshift relation \cite{Riess_04}
in CC \cite{Behnke_02,Behnke_04,zakhy}. Therefore,  the problem of nucleosynthesis
in the  Conformal Cosmology should be the subject of further study.

\vspace{.1cm}

{\bf 6. Large-scale structure in the Early Universe}.

The investigation of the large-scale structure in the Early Universe is one of the highlights of
the present-day Cosmology with far-reaching implications.

 In particular, the comparison of the cosmological perturbation theory in  the $\Lambda$CDM Model
 with the Hamiltonian approach to the same cosmological perturbation theory \cite{Barbashov_06}
 reveals essential differences of these approaches and their physical consequences.

 In order to demonstrate these consequences, we consider the case of integrable
  diffeo-invariant spacial coordinates, when the simplex components in interval (\ref{dg-4})
  ${\bf e}_{(b)i}dx^i={\omega}^{(3)}_{(b)}=dx_{(b)}$
 are total differentials.  The latter means
  that the coefficients of the spin-connection are equal to zero
  $
  \sigma_{(a)|(b)(c)}=
 {\bf e}_{(a)j}\left[\partial_{(b)}{\bf e}^j_{(c)}
 -\partial_{(c)}{\bf e}^j_{(b)}\right]=0
  $
  together with the  three-dimensional curvature $R^{(3)}=0$ in accord with observational data \cite{WMAP08}. In this case,  transverse gravitons cannot be oscillators and the renormalizablity problem
  should be reconsidered in both GR and DG.
  Eqs. (\ref{proi12}) and (\ref{proizvod1}) take the form
 \be\label{ls-1}
 p_{(b)(a)}=\frac{1}{3}v_{(ab)}=\frac{1}{6{\cal N}}\left(\frac{2}{3}\delta_{(a)(b)}\partial_{(c)}{\cal N}_{(c)}-\partial_{(a)}{\cal N}_{(b)}-\partial_{(b)}{\cal N}_{(a)}\right).
 \ee
 In this case, the transverse components of the shift vector can be defined by
  (\ref{h-c1})
 \bea\label{h-c1s}
 T_{(0)(a)}&=&
  -\partial_{(b)}p_{(b)(a)}+
 p_{\overline{\phi}}\partial_{(b)}\overline{\phi}+p_{\overline{Q}}\partial_{(b)}\overline{Q}=0,
  \eea
 while the shift vector longitudinal component   is defined by the constraint (\ref{1-43})
 $\partial_\tau e^{3\overline{D}}=
 \partial_{(b)}\left(e^{3\overline{D}}{\cal N}_{(b)}\right)$.
 The lapse function and dilaton are determined as solutions of Eqs.
 (\ref{3L+1G-8}), (\ref{3d}), (\ref{3L+1G-3}),  and
 (\ref{3L+1G-8d}).
  Solutions of these Eqs., in the first order in the Newton coupling constant,
  take forms \cite{Barbashov_06,B_06}
  \bea\label{12-17}
 e^{\overline{D}/2}
 &=&1+\frac{1}{2}\int d^3y\left[D_{(+)}(x,y)
\overline{T}_{(+)}^{(\mu)}(y)+
 D_{(-)}(x,y) \overline{T}^{(\mu)}_{(-)}(y)\right],\\\label{12-18}
 {\cal N}e^{7\overline{D}/2}
 &=&1-\frac{1}{2}\int d^3y\left[D_{(+)}(x,y)
\overline{T}^{(\nu)}_{(+)}(y)+
 D_{(-)}(x,y) \overline{T}^{(\nu)}_{(-)}(y)\right],
  \eea
 where
$D_{(\pm)}(x,y)$ are the Green functions satisfying
 the equations
 \bea\label{2-19}
 &&[\pm  m^2_{(\pm)}- \triangle
 ]D_{(\pm)}(x,y)=\delta^3(x-y),\\\label{2-19-1}
 &&m^2_{(\pm)}= 
 \dfrac{3(1+z)^2}{4}\left[{14(\beta\pm
1)}\Omega_{(0)}(z) \!\mp\!
 \Omega_{(1)}(z)\right]H_0^2,\\\label{2-19-2}
 &&\beta=\sqrt{\!1+\![\Omega_{(2)}(z)\!-\!\!14\Omega_{(1)}(z)
 ]/[98\Omega_{(0)}(z)]},\\\label{2-19-3}
 &&\Omega_{(n)}(z)=\sum\limits_{I=0,4,6,8,12}I^n(1+z)^{2-I/2}\Omega_{I},
 \eea
$\Omega_{I=0,4,6,8,12}$ are partial  density of states: rigid,
radiation, matter, curvature, $\Lambda$-term, respectively;
and $H_0$ is Hubble parameter,
 \bea\label{1cur1}\overline{T}^{(\mu)}_{(\pm)}
 &=&\overline{{\cal T}}_{(0)}\mp7\beta
  [7\overline{{\cal T}}_{(0)}-\overline{{\cal T}}_{(1)}],\\
  ~\overline{T}^{(\nu)}_{(\pm)}&=&[7\overline{\cal T}_{(0)}
 -\overline{\cal T}_{(1)}]
 \pm(14\beta)^{-1}\overline{\cal T}_{(0)}
 \eea
 are the local currents.

In the case of point mass distribution in a finite volume $V_0$ with the
zeroth pressure
  and  the  density
 \be\overline{{\cal T}}_{(0)}(x)=\dfrac{\overline{{\cal T}}_{(1)}(x)}{6}
  \equiv \dfrac{3}{4a^2} M\left[\delta^3(x-y)-\dfrac{1}{V_0}\right],\ee
 solutions   (\ref{12-17}),  (\ref{12-18}) take
 the Schwarzschild -type   form
 \bea\nonumber
  e^{\overline{D}/2}&=&1+
  \dfrac{r_{g}}{4r}\left[\dfrac{1+7\beta}{2}e^{-m_{(+)}(z)
 r}+ \dfrac{1-7\beta}{2}\cos{m_{(-)}(z)
 r}\right]_{H_0=0}=1+
  \dfrac{r_{g}}{4r},
 \\\nonumber
 {\cal N}e^{7\overline{D}/2}&=&1-
 \dfrac{r_{g}}{4r}\left[\dfrac{14\beta+1}{28\beta}e^{-m_{(+)}(z)
 r}+ \dfrac{14\beta-1}{28\beta}\cos{m_{(-)}(z)
 r}\right]_{H_0=0}=1-
 \dfrac{r_{g}}{4r},
 \eea
where $\beta=\sqrt{{25}/{49}}\simeq 1.01/\sqrt{2}$,~
$m_{(+)}=3m_{(-)}, m_{(-)}=H_0\sqrt{(1+z)\Omega_M \,3/2}
$.
These solutions have spatial oscillations and the
nonzero shift of the coordinate
  origin.

One can see that in the infinite volume limit $H_0=0,~a=1$
 these solutions  coincide with
 the isotropic version of  the Schwarzschild solutions in the  DG:
 $e^{\overline{D}/2}=1+\dfrac{r_g}{4r}$,~
 ${\cal N}e^{7\overline{D}/2}=1-\dfrac{r_g}{4r}$,~$N^k=0$.
 However, the analysis of the exact equations
 (\ref{3L+1G-8}), (\ref{3d}), (\ref{3L+1G-3}),  and
 (\ref{3L+1G-8d}) shows us that any nonzero cosmological density $\langle{T}^{1/2}_{\rm d}\rangle>0$ forbids negative values of the lapse function (\ref{3L+1G-8})
 ${\cal N}= {\langle{T}^{1/2}_{\rm d}\rangle}/
{{T}^{1/2}_{\rm d}}>0$ \cite{Barbashov_06,B_06,Glinka_08}.

 It is of interest  to find  an exact solution of these equations  for
 different equations of state.

There are the following differences of this relativistic perturbation theory  from the accepted
cosmological perturbation theory
for $e^{\overline{D}/2}=1-{\Psi}/{2}$,~
${\cal N}e^{3\overline{D}}=1+{\Phi}, N_k=0, h_{ij}=0$ and ${\cal H}=a'/a$ \cite{lif,mukh}.
     The $\Lambda$CDM Model treats the scalar metric component as a dynamic variable,
        omits the decomposition of
 the potential part of the energy density, that leads to  spacial oscillations and the large scale
 structure of the Universe,  chooses the gauge with the zero shift vector,
and destroys the  Hamiltonian approach by the double counting of the zeroth Fourier-harmonics of the scalar metric component \cite{Barbashov_06}.

 {\bf 7.  Luminosity-distance -- redshift relation
in CC.}

Assuming that supernovae type Ia (SNe Ia) are standard candles one
could use them to test cosmological theories. The Hubble Space
Telescope team analyzed  186 SNe Ia\cite{Riess_04} to test the
Standard Cosmological model (SC) associated with expanded lengths
in the Universe and evaluate its parameters. The problem is to use the same
sample to determine parameters of Conformal Cosmological model \cite{Behnke_02,Behnke_04,zakhy}.

\section{Conclusion}

After realization of this program one can expect
 that DG and  the Standard Model
  supplemented with a free scalar field  $Q$ in the CMB
 reference frame with the initial data (\ref{in-d1})
do not contradict the following scenario of the evolution of the
Universe within conformal cosmology \cite{Blaschke_04}:\\
 $\eta \sim 10^{-12}s,$ {creation of vector bosons from a
 ``vacuum''};\\ [1.5mm] $10^{-12}s < \eta <
10^{-11}\div 10^{-10} s,$ {formation of baryon-antibaryon asymmetry;}\\
 $\eta \sim 10^{-10}s,$ {decay of vector bosons;}\\
 $10^{-10}s <\eta < 10^{11}s,$ { primordial chemical
evolution of matter;}\\ 
$\eta \sim 10^{11}s,$ {recombination
or separation of cosmic
microwave background radiation;}\\
 $\eta \sim  10^{15}s,$ {formation of galaxies;}\\
  $\eta > 10^{17}s,$ { terrestrial experiments and evolution
of supernovae.}

 The relativistic physics gives us the Foundations of  Relativistic Cosmology of the Universe
 as the historically existing system of classification of observational and experimental facts
 in physics and astrophysics.

  There are tendencies of modern cosmology to ignore these principles,  in particular
 to  replace the initial data by the fundamental parameters of the motion equations.
 However, this replacement  leads to contradictions
 in acceptable modern models in applying
 mathematical tools of the type of the relativistic classification of physical states, or
 the Hamiltonian method because the latter
   were developed
 especially to solve equations with the initial data.

  Nevertheless, ignoring relativistic classification of physical states  and initial data,
  and confusing the gauge transformations with the  frame ones \cite{lif}, modern cosmological models
  \cite{Giovannini,linde,mukh}
  treat  the CMB  motion with the velocity
  368 km/s to Leo with respect to  the frame of reference of the Earth observer
  as object of relativistic transformations of the type of a change of the frame.
      This change of the frame leads to a change of the measurable parameters of the CMB temperature dipole component, i.e. initial data.

  If the CMB and the Universe  are gauge-invariant objects of the relativistic
  transformations, this means that the Universe as a whole is an object of relativistic physics
  as a theory of initial data,  including Poincar\'e
 group, its Wigner representations in the form of the frame-covariant and gauge-invariant physical states,
 the N\"other theorem gauge constraints, their consequences as
 derivation of the minimal field theories of type of GR and SM, and gauge-invariant solutions of the
 field equations supplemented by the  cosmological Initial Data given at the beginning, when there were nothing
 but homogeneous scalar fields.  What is  Origin of these Initial Data? What are a group
 of their transformations, irreducible unitary representations, the N\"other theorem gauge constraints,
 ''field theories'', and their physical consequences?

%

\end{document}